%% file: Template.tex
\documentclass{article}
\usepackage{spconf,amsmath,graphicx,}
\usepackage{acronym}
\usepackage{svg}
\usepackage{cite}
\acrodef{SSSR}{self-supervised speech representation}
\acrodef{CNN}{convolutional neural network}
\acrodef{MSE}{mean squared error}
\acrodef{MOS}{mean opinion score}
\def\set@curr@file#1{\def\@curr@file{#1}}

\input{will_acronym.tex}

\title{Perceive and predict: self-supervised speech representation based loss functions for speech enhancement} 
\name{George Close, William Ravenscroft, Thomas Hain and Stefan Goetze\thanks{This work was supported by the Centre for Doctoral Training in Speech and Language Technologies (SLT) and their Applications funded by UK Research and Innovation [grant number EP/S023062/1]. This work was also funded in part by TOSHIBA Cambridge Research Laboratory and 3M Health Information Systems Inc.}}
\address{Department of Computer Science, University of Sheffield}
\begin{document}
\ninept
\maketitle
\begin{abstract}
Recent work in the domain of speech enhancement has explored the use of self-supervised speech representations to aid in the training of neural speech enhancement models.
However, much of this work focuses on using the deepest or final outputs of self supervised speech representation models, rather than the earlier feature encodings. The use of self supervised representations
in such a way is often not fully motivated. 
In this work it is shown that the distance between the feature encodings of clean and noisy speech correlate strongly with psychoacoustically motivated measures of speech quality and intelligibility, as well as with human \ac{MOS} ratings.
Experiments using this distance as a loss function are performed and improved performance over the use of STFT spectrogram distance based loss as well as other common loss functions from speech enhancement literature is demonstrated using objective measures such as \ac{PESQ} and \ac{STOI}.
\end{abstract}
\section{Introduction}
\label{sec:intro}
Speech enhancement remains an active area of speech research due to its applications in numerous downstream tasks \cite{FFASRHaebUmbach}. Deep learning models have led to state-of-the-art results on numerous benchmarks for speech enhancement and related tasks \cite{fu2021metricgan,WHAMR,mimospeech,close2022}. A key area of research for these kinds of models has been the loss functions used \cite{selosses}. Furthermore,  metrics used to evaluate speech enhancement models have long been an active area of research; in many cases they can also be used as the loss function used to train the models themselves \cite{PESQ,STOI,LeRoux,fu2021metricganu}.
\Acp{SSSR} are of increasing interest in a number of speech related tasks, including speech enhancement \cite{wav2vec,hubert,superb}.
Generally, the strength of \ac{SSSR} comes from their ability to predict the context of the speech content in the input audio, and thus model the patterns of spoken language.
In this work, it is investigated whether \ac{SSSR} distances between clean and noisy speech signals have a stronger correlation to perceptually motivated speech enhancement measures such as \ac{PESQ} than the spectral distances of the signals. This analysis is also repeated for \ac{MOS}. In the contribution here we consider both the output of \ac{SSSR} encoder layers as well as the final output layer. It is demonstrated the encoder layers have a notably stronger correlation to the aforementioned evaluation measures than the output layers.
\Ac{HuBERT}~\cite{hubert} and XLSR~\cite{xlsr} \acp{SSSR} are chosen as these have both been applied in related speech tasks previously but in different ways \cite{meta_phone_aware_se,nisqa_pretrained_ss}.
Following from this, the distances between clean and noisy \ac{SSSR} features are then evaluated for their usefulness as loss functions to train speech enhancement models. 
Their performance is compared to other standard loss functions including perceptually motivated loss functions such as \ac{STOI} loss \cite{stoi_loss}.

The remainder of this paper is structured as follows. In Section~\ref{sec:sssr_intro}, \acp{SSSR} are introduced and the two specific models used in this work are detailed. In Section~\ref{sec:metric_comparison} the relationship between distance measures derived from these models with speech assessment metrics and human assessment is analysed, in order to obtain novel insight in to the use of \ac{SSSR}s for perceptually motivated speech enhancement. Finally, in Section~\ref{sec:experiment} simple signal enhancement networks are trained using loss functions derived from \ac{SSSR} distances and are compared to baselines using conventional loss functions, analysing the distance measures usefulness as loss function for a single channel speech enhancement task.
\section{Self Supervised Speech Representation Models}\label{sec:sssr_intro}
\ac{SSSR} models are neural models of speech which are trained in a self supervised way. This is typically done by `masking' a portion of the input and then tasking the model with recreating the masked portion. At inference time, the network layers responsible for the recreation step are removed and the model instead returns a deep `context' representation of the input time domain audio. At this point, additional task specific layers can be appended to the network, with the self supervised representation model either being fine-tuned or frozen as the task specific layers are trained. Generally speaking, \ac{SSSR}s can be said to first \textit{perceive} the input audio in a feature encoder step, and then \textit{predict} the context of the content of the audio in the deeper layers. 
\\
\textbf{HuBERT}~\cite{hubert} (Hidden Unit BERT) is a \ac{SSSR} model which utilises a BERT \cite{bert} style prediction loss in training. 
It consists of two main components; the first is  a \ac{CNN} feature encoder block consisting of stacked 1 dimensional convolutional layers which transform the input time domain audio signal 
into a 512 channel feature representation. This is followed by a transformer~\cite{transformer} block which consists of several transformer layers to produce the final 768 channel output representation.
The HuBERT model used in this work is trained on the 960h Librispeech~\cite{7178964} training set and is sourced from the fairseq Github repository\footnote{\texttt{https://github.com/facebookresearch/fairseq}}. 
\\
The \textbf{Cross Lingual Speech Representation (XLSR)} \cite{xlsr} model is a variant of the Wav2Vec2~\cite{wav2vec} \ac{SSSR} model. It is trained on 436,000 hours of speech data from a number of different languages, including BABEL\footnote{\texttt{https://catalog.ldc.upenn.edu/byyear}}, which contains potentially noisy telephone conversations in a number of languages. It is structured similarly to HuBERT, with a convolutional feature encoder block which outputs a 512 channel feature representation, followed by a transformer block which outputs a final representation with 1024 channels. In this work we make use of the smallest version of XLSR, sourced from the official HuggingFace repository\footnote{\texttt{https://huggingface.co/facebook\\/wav2vec2-xls-r-300m}}. 
\subsection{\ac{SSSR}s in quality estimation and speech enhancement}
In \cite{nisqa_pretrained_ss}, a non intrusive human \ac{MOS} predictor is proposed which makes use of pretrained XLSR representations as a feature extraction stage. With only a simple trained prediction network placed over these XLSR representations, the proposed model was able to achieve high performance in the Conferencing Speech 2022~\cite{cs2022quality} quality prediction challenge. This indicates that the XLSR \ac{SSSR} model is able to capture quality information about the input signal, even without access to a reference. It's training data differs to that of other \ac{SSSR} models (such as HuBERT) in that it is trained in part on non-clean speech. In this work, we aim to analyse the behaviour of XLSR and HuBERT when exposed to noisy/low quality audio. 
\\
In \cite{meta_phone_aware_se}, a number of techniques to incorporate \ac{SSSR}s (namely HuBERT) into a single channel speech enhancement system are proposed. One of these techniques,  called `supervision' in \cite{meta_phone_aware_se} involves the use of the distance between the \ac{SSSR} output representations of clean reference speech and the enhanced noisy speech output by the enhancement model as an additional loss term to train the model. This is in turn inspired by a prior work~\cite{perceptual_quality_phone_fort} in which the Wasserstein distance between clean and enhanced \ac{SSSR} representations of the audio is used as a loss term. In \cite{perceptual_quality_phone_fort} the relationship between the \ac{SSSR} distances and perceptual measures is noted. However, both of these works are motivated by the phoneme level information encoded in later layers of the \ac{SSSR} network and thus only consider the distance between the clean and noisy representations of the \textit{final output layer} of \ac{SSSR} models. Here, we aim to explore the use of distances derived from the intermediate \textit{feature encoder layer} of the \ac{SSSR} models. Moreover, we aim to directly compare \ac{SSSR} derived distances with a standard distance used as a loss function for speech enhancement, by formulating the proposed distance measures similarly, i.e. as \ac{MSE} distances.  

\section{\ac{SSSR} derived distances in relation to speech assessment metrics}\label{sec:metric_comparison}
In this work, first the \acf{MSE} distance between representations of some clean speech $s[n]$ and a corresponding noisy version of $s[n]$, $x[n]$
\label{signal_model}
\begin{equation}
    x[n] = s[n] + v[n],
    \label{eq:basic}
\end{equation}
is analysed, where $n$ is the discrete time index and $v[n]$ is some additive noise caused by the recording environment. Specifically, we define these using either $\mathbf{S}_\mathrm{FE}$, $\mathbf{X}_\mathrm{FE}$ or $\mathbf{S}_\mathrm{OL}$, $\mathbf{X}_\mathrm{OL}$ where $\mathrm{FE}$ and $\mathrm{OL}$ denote the \ac{SSSR} encoder representation and final output layer respectively: 
\begin{equation}
    \label{eq:fe_transform}
    \mathbf{S}_\mathrm{FE} = \mathcal{G}_\mathrm{FE}(s[n])
\end{equation}
\begin{equation}
    \label{eq:ol_transform}
    \mathbf{S}_\mathrm{OL} = \mathcal{G}_\mathrm{OL}(\mathcal{G}_\mathrm{FE}(s[n]))
\end{equation}
with  $\mathcal{G}_\mathrm{FE},\mathcal{G}_\mathrm{OL}$ denoting the encoder output and final layer in the \ac{SSSR} mode respectively, and $\mathbf{X}_\mathrm{FE}$, $\mathbf{X}_\mathrm{OL}$ defined similarly. $\mathbf{S}_\mathrm{FE}$ is the output of the \ac{SSSR}s feature encoder layer, while $\mathbf{S}_\mathrm{OL}$ is the output of its final layer.
The \ac{MSE} distances between these \ac{SSSR}  representations are defined as:
\begin{equation}
    \label{eq:enc_mse}
    d_{\mathrm{FE}}(\mathbf{S}_\mathrm{FE},\mathbf{X}_\mathrm{FE}) = \sum_{t,f}^{T,F} (\mathbf{S}_{\mathrm{FE}}[t,f] - \mathbf{X}_{\mathrm{FE}}[t,f])^2 
\end{equation}

\begin{equation}
    \label{eq:out_mse}
    d_{\mathrm{OL}}(\mathbf{S}_\mathrm{OL},\mathbf{X}_\mathrm{OL}) = \sum_{t,f}^{T,F} (\mathbf{S}_{\mathrm{OL}}[t,f] - \mathbf{X}_{\mathrm{OL}}[t,f])^2 
\end{equation}
where $T$ and $F$ denote time and feature dimensions of the representation.


These \ac{SSSR} derived distances are compared with a distance which is commonly used as a loss function in speech enhancement tasks. The \ac{MSE} distance between clean and noisy spectrogram representations is taken:
\begin{equation}
    \label{eq:spec_mse}
    d_{\mathrm{SG}}(\mathbf{S}_\mathrm{SG},\mathbf{X}_\mathrm{SG}) = \sum_{t,f_{Hz}}^{T,F_{Hz}} \left(\mathbf{S}_{\mathrm{SG}}[t,f_{Hz}] - \mathbf{X}_{\mathrm{SG}}[t,f_{Hz}]\right)^2 
\end{equation}
where $\mathbf{S}_\mathrm{SG}$ and $\mathbf{X}_\mathrm{SG}$ are magnitude spectrogram representations of $s[n]$ and $x[n]$, respectively, and $T$ and $F_{Hz}$ are the time and frequency dimensions of the spectrograms.\\ In the following, $d_{\mathrm{FE}}$ and $d_{\mathrm{OL}}$ are computed using the XLSR and HuBERT models. As mentioned in the previous section, $\mathbf{S}_\mathrm{FE}$, $\mathbf{X}_\mathrm{FE}$ and $\mathbf{S}_\mathrm{OL}$, $\mathbf{X}_\mathrm{OL}$ of the XLSR model have feature dimensions $F$ of $512$ and $1024$ respectively, while those of HuBERT have feature dimensions of $512$ and $768$, sharing a time dimension $T$, the size of which is dependent on the length in samples of the input time domain audio.\\ $\mathbf{S}_\mathrm{SG}$ and $\mathbf{X}_\mathrm{SG}$ are computed using a Fourier Transform with an FFT size of $512$, window length of $32$ms, and a hop length of $16$ms (resulting in a 50\% frame overlap) using a hamming window. This results in a spectogram with a frequency dimension $F_{Hz}$ of $257$, and a time dimension $T$ which is dependent on the length in samples of the input time domain audio. 
%
\subsection{Datasets}
To express the relationship between the distance measures and psychoacoustically motivated metrics the VoiceBank-DEMAND \cite{vb-demand}, a popular and commonly used dataset for single channel speech enhancement, is used.
Its training set consists of $11572$ clean and noisy speech audio file pairs  $(s[n], x[n])$, mixed at four different \acp{SNR} \{$0$, $5$, $10$, $15$\}~dB. Eight noise files are sourced from the DEMAND~\cite{demand} noise dataset; two others, a babble noise and a speech-shaped noise were also used.
The training set contains speech from 28 different speakers ($14$ male, $14$ female), with English or Scottish accents.The testset containing $824$ utterances is mixed at SNRs of {$2.5$, $7.5$, $12.5$ and $17.5$}~dB, with five different noises which do not appear in the training set from the DEMAND corpus.
\\
In order to assess the relationship between the distance measures and human \ac{MOS} ratings, the NISQA~\cite{NISQA} dataset is used. This is a dataset of variable length clean and noisy speech audio file pairs $(s[n], x[n])$ with real human-annotated \ac{MOS} labels, designed for the training and testing of neural \ac{MOS} predictors.The two testsets used here contain 440 clean/noisy pairs in total. \\ 
The audio files in both datasets have a sample rate of $48$ kHz and are down-sampled to $16$ kHz such that $\mathcal{G}(\cdot)$ in (\ref{eq:fe_transform}), (\ref{eq:ol_transform}) can computed.

\subsection{\ac{SSSR} distances and psychoacusitcally motivated metrics}
\begin{figure}
\begin{minipage}{1.0\linewidth}
  \centering
  \centerline{\includegraphics[width=8.5cm]{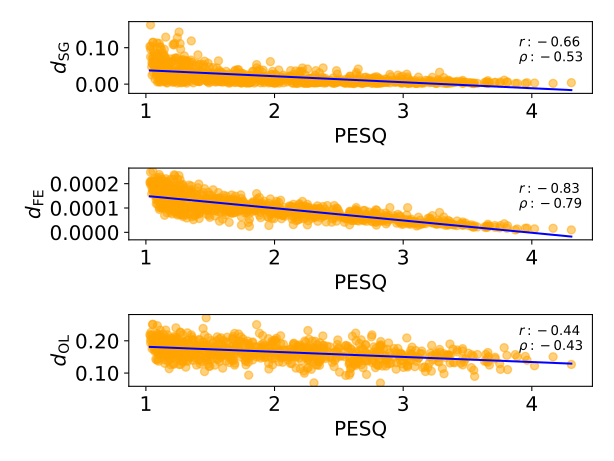}}
  \caption{Scatter plots showing the relationship between  the PESQ metric and \ac{MSE} Spectrogram distance   $d_\mathrm{SG}$ as well as, HuBERT $d_\mathrm{FE}$, HuBERT  $d_\mathrm{OL}$ distances for the VoiceBank-DEMAND testset}\label{fig:pesq_scatters}
\end{minipage}
\end{figure}

\begin{table*}
\centering
\begin{tabular}{c|cc|cc|cc|cc|cc|cc}
\centering
\textbf{Distance} & \multicolumn{2}{c|}{\textbf{PESQ}}                     & \multicolumn{2}{c|}{\textbf{STOI}}                     & \multicolumn{2}{c|}{\textbf{Csig}}                     & \multicolumn{2}{c|}{\textbf{Cbak}}                     & \multicolumn{2}{c|}{\textbf{Covl}} & \multicolumn{2}{c}{\textbf{MOS}}                      \\ \hline
                 & \multicolumn{1}{c|}{$r$}           & $\rho$            & \multicolumn{1}{c|}{$r$}           & $\rho$            & \multicolumn{1}{c|}{$r$}           & $\rho$            & \multicolumn{1}{c|}{$r$}           & $\rho$            & \multicolumn{1}{c|}{$r$}           & $\rho$      & \multicolumn{1}{c|}{$r$}           & $\rho$      \\
 $d_\mathrm{SG}$          & \multicolumn{1}{l|}{-0.66}          & -0.53       & \multicolumn{1}{l|}{-0.60}          & -0.68         & \multicolumn{1}{l|}{-0.75}          & -0.70         & \multicolumn{1}{l|}{-0.84}          & -0.69        & \multicolumn{1}{l|}{-0.74}          & -0.64  & \multicolumn{1}{l|}{0.35}           & -0.27          \\
XLSR $d_\mathrm{FE}$           & \multicolumn{1}{l|}{-0.82} & -0.78 & \multicolumn{1}{l|}{\textbf{-0.80}} & \textbf{-0.81} & \multicolumn{1}{l|}{-0.93} & -0.92 & \multicolumn{1}{l|}{-0.88} & -0.85 & \multicolumn{1}{l|}{-0.90} & -0.87  & \multicolumn{1}{l|}{-0.47} & -0.43\\
XLSR $d_\mathrm{OL}$          & \multicolumn{1}{l|}{-0.66}          & -0.61          & \multicolumn{1}{l|}{-0.69}          & -0.68          & \multicolumn{1}{l|}{-0.76}          & -0.75          & \multicolumn{1}{l|}{-0.74}          & -0.72          & \multicolumn{1}{l|}{-0.74}          & -0.71 & \multicolumn{1}{l|}{-0.44}          & -0.40 \\
HuBERT $d_\mathrm{FE}$          & \multicolumn{1}{l|}{\textbf{-0.83}}          & \textbf{-0.79}         & \multicolumn{1}{l|}{-0.75}          & -0.76          & \multicolumn{1}{l|}{\textbf{-0.95}}          & \textbf{-0.93}          & \multicolumn{1}{l|}{\textbf{-0.90}}         & \textbf{-0.87}          & \multicolumn{1}{l|}{\textbf{-0.91}}          & \textbf{-0.89} &\multicolumn{1}{l|}{\textbf{-0.48}} &\textbf{-0.46}\\
HuBERT $d_\mathrm{OL}$          & \multicolumn{1}{l|}{-0.44}          & -0.43          & \multicolumn{1}{l|}{-0.40}          & -0.42          & \multicolumn{1}{l|}{-0.52}          & -0.52          & \multicolumn{1}{l|}{-0.45}          & -0.45          & \multicolumn{1}{l|}{-0.50}          & -0.49 & \multicolumn{1}{l|}{-0.42} &-0.37  
\end{tabular}
\caption{Spearman $r$ and Pearson $\rho$ correlation between distance measures and speech quality and intelligibility metrics in the VoiceBank-DEMAND testset, as well as \ac{MOS} in the NISQA Challenge testset }\label{tab:metric_corrs}
\end{table*}
Fig.~\ref{fig:pesq_scatters} shows the relationships between the  distance measures and PESQ~\cite{PESQ} scores computed using the $s[n]$, $x[n]$ pairs in the VoiceBank-DEMAND~\cite{vb-demand} testset using HuBERT \ac{SSSR}.  From these, it can be observed that both $d_{\mathrm{FE}}$ and $d_{\mathrm{OL}}$ correlate significantly more strongly than $d_{\mathrm{SG}}$ with quality  metrics. Furthermore, the distance computed using the output of the 1D convolutional block $d_{\mathrm{FE}}$ correlates more strongly than the distance computed using the \ac{SSSR} output $d_{\mathrm{OL}}$. This suggests that the phonetic and linguistic processing which occurs in the deeper parts of the  model are less sensitive to the noise in $x[n]$.
The first 5 columns of Table~\ref{tab:metric_corrs} shows the Spearman $r$ and Pearson correlations $\rho$ between PESQ, STOI and the components of the Composite~\cite{composite} measure. Like with \ac{PESQ},\ac{STOI} and the Composite measure scores all correlate more strongly with the proposed \ac{SSSR} distances than with $d_\mathrm{SG}$. To our knowledge, this is the first time that the correlation between \ac{SSSR} derived distances and the Composite measure have been analysed, and the  high correlation displayed here shown. 

\subsection{\ac{SSSR} distances and human quality assessment}
\begin{figure}
\begin{minipage}{1.0\linewidth}
  \centering
  \centerline{\includegraphics[width=8.5cm]{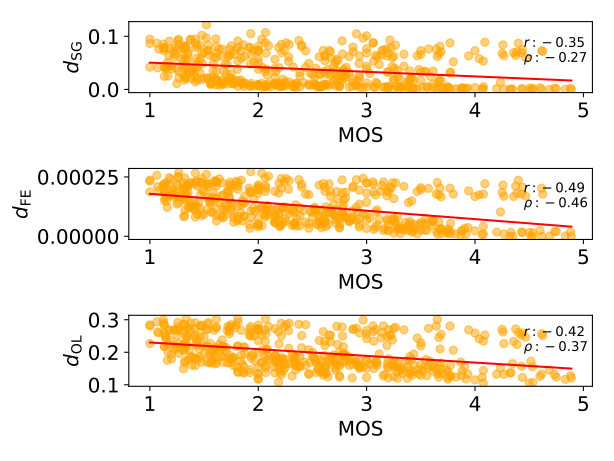}}
  \caption{Scatter plots showing the relationship between human MOS scores and \ac{MSE} Spectrogram distance $d_\mathrm{SG}$, HuBERT $d_\mathrm{FE}$ and HuBERT $d_\mathrm{OL}$ with  in the NISQA Challenge testset}\label{fig:mos_scatters}
\end{minipage}
\end{figure}

Fig.~\ref{fig:mos_scatters} and  the last column of Table~\ref{tab:metric_corrs} show the relationship between the MSE distances and  human \acp{MOS} in the `FOR' and `P501' testset  $s[n]$, $x[n]$ pairs of the NISQA~\cite{NISQA} dataset.
While the overall correlations are lower here than those of the metrics analysed in the first 5 colunms, the same pattern emerges with  $d_\mathrm{FE}$ and $d_\mathrm{OL}$ correlating more strongly with the \ac{MOS} scores than $d_\mathrm{SG}$. The HuBERT based distances again correlate more strongly than XLSR; this is possibly due to the language match between the training data of HuBERT and that of the data, both being English only. 


\section{\ac{SSSR} Based Signal Enhancment Experiment}\label{sec:experiment}
An experiment is carried out in order to assess the effectiveness of the \ac{SSSR} derived distance measures as loss functions for speech enhancement tasks.
\subsection{Experiment setup}
 Simple masking based speech enhancement models were trained using a number of different loss functions; $d_\mathrm{SG}$, $d_\mathrm{FE}$, $d_\mathrm{OL}$ as described in the previous sections, as well as Si-SDR loss~\cite{tasnet} and STOI loss~\cite{stoi_loss}. 
 Each model was trained for 50 epochs on the VoiceBank-DEMAND~\cite{vb-demand} training set. $d_\mathrm{FE}$, $d_\mathrm{OL}$ are computed for XLSR or HuBERT feature encoder and output representations. The Adam~\cite{kingma2017adam} optimiser is used with a learning rate of $0.001$.  At test time, the epoch obtaining the highest \ac{PESQ} score on the validation set is loaded. The SpeechBrain~\cite{speechbrain} toolkit is used to implement the experiment. 
\subsection{Loss Functions}
The distance measures defined in (\ref{eq:enc_mse}), (\ref{eq:out_mse}) and (\ref{eq:spec_mse}) are modified to be used as loss terms for a speech enhancement neural model:
\begin{equation}
    \label{eq:enc_loss}
    L_{\mathrm{FE}}(\mathbf{S}_\mathrm{FE},\mathbf{\hat{S}}_\mathrm{FE}) = \sum_{t,f}^{T,F} (\mathbf{S}_{\mathrm{FE}}[t,f] - \mathbf{\hat{S}}_{\mathrm{FE}}[t,f])^2 
\end{equation}
\begin{equation}
    \label{eq:out_loss}
    L_{\mathrm{OL}}(\mathbf{S}_\mathrm{OL},\mathbf{\hat{S}}_\mathrm{OL}) = \sum_{t,f}^{T,F} (\mathbf{S}_{\mathrm{OL}}[t,f] - \mathbf{\hat{S}}_{\mathrm{OL}}[t,f])^2 
\end{equation}
\begin{equation}
    \label{eq:spec_loss}
    L_{\mathrm{SG}}(\mathbf{S}_\mathrm{SG},\mathbf{\hat{S}}_\mathrm{SG}) =  \sum_{t,f_{Hz}}^{T,F_{Hz}} (\mathbf{S}_{\mathrm{SG}}[t,f_{Hz}] - \mathbf{\hat{S}}_{\mathrm{SG}}[t,f_{Hz}])^2 
\end{equation}
where $\hat{s}[n]$ is the enhanced time domain audio signal output by the neural model when $x[n]$ is input and $\mathbf{\hat{S}}_\mathrm{FE}$, $\mathbf{\hat{S}}_\mathrm{OL}$ and $\mathbf{\hat{S}}_\mathrm{SG}$ are the feature encoder output, output layer and spectrogram representations of $\hat{s}[n]$ respectively.
\subsection{Enhancement Model Structure}
The same model structure is used for all models. It consists of 2 \ac{BLSTM} layers followed by two linear layers, with the first linear layer using a LeakyReLU activation and the second a Sigmoid. The input to the model is a magnitude spectrogram ($\mathbf{X}_\mathrm{SG}$) and the model returns a `mask' which is multiplied with the noisy input spectrogram to produce the enhanced spectrogram $\mathbf{\hat{S}}_\mathrm{SG}$. From this, the enhanced time domain speech $\hat{s}[n]$ is then created via the overlap-add resynthesis method, using the original noisy phase of $x[n]$. This structure is selected because, despite being relatively simple and with a small parameter count, it is able to achieve state of the art performance in perceptually motivated speech enhancement~\cite{fu2021metricgan,fu2021metricganu,close2022}.
%
\vspace{-0.5cm}
\subsection{Signal Enhancement Performance}
Table~\ref{tab:results1} shows the experiment results. The proposed model using HuBERT $L_\mathrm{FE}$ as its loss function outperforms the baseline using the spectrogram distance $L_\mathrm{SG}$ in terms of \ac{PESQ} and the Composite measure Csig, Cbak and Covl. Additionally, the best performing model by a significant margin in terms of Cbak uses the XLSR encoder distance loss function $L_\mathrm{FE}$, and most \ac{SSSR} based losses outperform the baseline systems in this measure. Those models which use \ac{SSSR} encoder distance $L_\mathrm{FE}$ outperform those which use \ac{SSSR} output layer distance $L_\mathrm{OL}$; this is consistent with the correlation values in Table~\ref{tab:metric_corrs} where $d_\mathrm{FE}$ distances correlate more strongly with the metrics than $d_\mathrm{OL}$ distances. 
\begin{table}[!htb]
\centering
\resizebox{\columnwidth}{!}{
\begin{tabular}{l|rrrrrrr}
\textbf{loss function} & \multicolumn{1}{c}{\textbf{PESQ}} & \multicolumn{1}{c}{\textbf{STOI}} & \multicolumn{1}{c}{\textbf{Csig}} & \multicolumn{1}{c}{\textbf{Cbak}} & \multicolumn{1}{c}{\textbf{Covl}} & \multicolumn{1}{c}{\textbf{Si-SDR}} \\ \hline
\textit{noisy}              & \textit{1.97}                     & \textit{0.92}                     & \textit{3.35}                     & \textit{2.44}                     & \textit{2.63}                     & \textit{8.98}                                       \\
$d_\mathrm{SG}$            & 2.70                     & \textbf{0.94}                     & 4.00                     & 2.62                              & 3.35                     & 18.62     \\
${L}_\mathrm{SISDR}$~\cite{tasnet}         & 2.28          & 0.92          & 3.51           & 2.44           & 2.88           & \textbf{18.66}                       \\
 $L_{\mathrm{STOI}}$~\cite{stoi_loss}        & 2.12          & 0.93          & 3.46           & 2.16          & 2.77           & 13.31                  \\ \hline
HuBERT $L_\mathrm{FE}$               & \textbf{2.79}                     & 0.94                              & \textbf{4.10}                     & 2.68                              & \textbf{3.44}                     & 18.47                                                       \\
HuBERT $L_\mathrm{OL}$                & 2.55                              & 0.92                              & 3.66                              & 2.42                              & 3.08                              & 14.92                                                          \\
XLSR $L_\mathrm{FE}$                 & 2.69                              & 0.92                              & 3.77                              & \textbf{3.05}                     & 3.21                              & 9.72                                                   \\
XLSR $L_\mathrm{OL}$                  & 2.43                              & 0.91                              & 3.21                              & 2.64                              & 2.79                              & 13.00                                                         
\end{tabular}
}
\caption{Signal Enhancement performance on the VoiceBank-DEMAND testset}\label{tab:results1}
\end{table}
\vspace{-0.5cm}
\subsection{Analysis}
\begin{figure}[!htb]
  \centering
  \centerline{
  \includegraphics[width=\columnwidth]{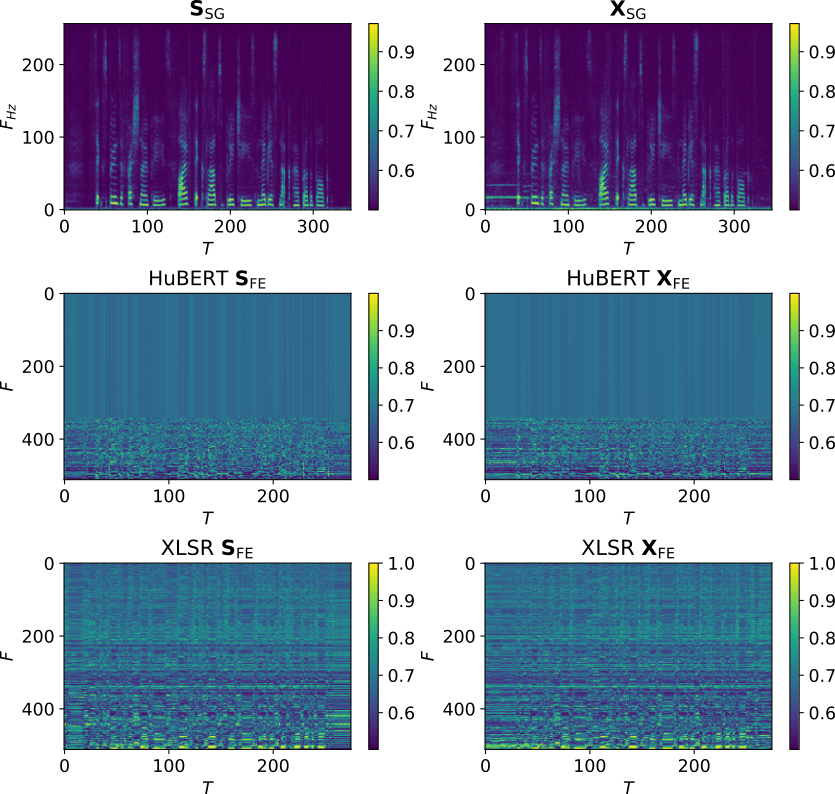}
  }
  \caption{Visualisation of inputs representations of $s[n]$, $x[n]$ to $d_\mathrm{SG}$, HuBERT $d_\mathrm{EF}$ and XLSR $d_\mathrm{EF}$. \Ac{SSSR} features are sorted according to depthwise euclidean distance following Algorithm 1 in \cite{atttasnet} and a sigmoid function is applied to increase clarity.}\label{fig:feat_reps}
\end{figure}

Fig.~\ref{fig:feat_reps} shows an example of the feature representations of $s[n]$ and $x[n]$ used as inputs to $d_\mathrm{SG}$ and $d_\mathrm{FE}$ for both HuBERT and XLSR.
Tonal noise introduced in $x[n]$, visible as a line spanning approximately the first 50  time frames in $\mathbf{X}_\mathrm{SG}$, is well represented in the XLSR $\mathbf{X}_\mathrm{FE}$ but not in HuBERT $\mathbf{X}_\mathrm{FE}$. This is a possible explanation for the increased Cbak score for the XLSR $L_{FE}$ loss over the HubERT  based loss $L_{FE}$ as XLSR $\mathbf{X}_\mathrm{FE}$ representations appear to be more sensitive to noise in non speech regions of the representation. The fact that XLSR is trained in part on noisy data is a potential explanation for this  behaviour. 

\section{Conclusion}
In this work it is demonstrated that the earlier `perceive' feature encoder layers of \acp{SSSR} preserve aspects of noise and distortion in speech to a greater degree than the deeper `predict' layers. Moreover, we find that a simple distance measure between the encoder representations of clean and noisy speech correlates strongly with perceptually motivated metrics of speech quality, as well as with human speech quality assessment. This correlation is affected by the attributes of the data used to train the \ac{SSSR}. This finding is validated by the use of these distance measures as loss functions for a speech enhancement task, where feature encoder distance outperforms both the deeper output layer and a standard spectrogram based loss. Future work will include the further tuning of \ac{SSSR} encoders towards human perception, as well as investigating the effect of the training data on the \ac{SSSR} encoder representations.  
\bibliographystyle{IEEEbib}
\bibliography{IEEEfull,will}
\end{document}

%% file: will_acronym.tex
\acrodef{AC}    {acoustic conditions}
\acrodef{ASR}   {automatic speech recognition}
\acrodef{BERT}  {bidirectional encoder representations from transformers}
\acrodef{BLSTM} {bidirectional long short term memory}
\acrodef{CB}    {convolutional block}
\acrodefplural{CB}[CBs]{convolutional blocks}
\acrodef{Conv-TasNet} {convolutional \ac{TasNet}}
\acrodef{CSM}   {clean speech mixtures}
\acrodef{cLN}   {channelwise layer normalization}
\acrodef{DAE}   {denoising autoencoder}
\acrodef{D-Conv}{depthwise convolution}
\acrodef{DD-Conv}{deformable depthwise convolution}
\acrodef{DDS-Conv}{deformable depthwise-separable convolution}
\acrodef{DS-Conv}{depthwise-separable convolution}
\acrodef{DL}    {deep learning}
\acrodef{DNN}   {deep neural network}
\acrodef{DPRNN} {dual path recurrent neural network model}
\acrodef{DPTNet}{dual path Transformer network}
\acrodef{DTCN}  {Deformable Temporal Convolutional Network}
\acrodef{E2E}   {end-to-end}
\acrodef{ER}    {early reflection}
\acrodefplural{ER}[ERs]{early reflections}
\acrodef{gLN}   {global layer normalization}
\acrodef{HuBERT}{hidden unit BERT}
\acrodef{LSTM}  {long short term memory}
\acrodef{LR}    {late reflection}
\acrodefplural{LR}[LRs]{late reflections}
\acrodef{MHA}   {Multihead Attention}
\acrodef{MOS}   {Mean Opinion Score}
\acrodef{MR}    {mask refinement}
\acrodef{NMF}   {non-negative matrix factorization}
\acrodef{NSM}   {noisy speech mixture}
\acrodef{NRSM}  {noisy reverberant speech mixture}
\acrodef{PESQ}  {perceptual evaluation of speech quality}
\acrodef{PIT}   {permutation invariant training}
\acrodef{PM}    {post-masking}
\acrodef{P-Conv}{pointwise convolution}
\acrodef{PReLU} {parametric \ac{ReLU}}
\acrodef{ReLU}  {rectified linear unit}
\acrodef{RF}    {receptive field}
\acrodef{RIR}   {room impulse response}
\acrodef{RIRs}   {room impulse responses}
\acrodefplural{RIR}[RIRs]{room impulse responses}
\acrodef{RSM}   {reverberant speech mixture}
\acrodef{SISDR} {scale-invariant signal-to-distortion ratio}
\acrodef{SDR}   {signal-to-distortion ratio}
\acrodef{SNR}   {signal-to-noise ratio}
\acrodef{SRMR}  {speech-to-reverberation modulation energy ratio}
\acrodef{SP}    {signal processing}
\acrodef{STFT}  {short-time Fourier transform}
\acrodef{STOI}  {short-time objective intelligibility}
\acrodef{SOTA}  {state-of-the-art}
\acrodef{ESTOI} {extended short-time objective intelligibility}
\acrodef{TasNet}{Time-domain audio separation network}
\acrodef{TCN}   {temporal convolutional network}
\acrodef{UPGMA} {unweighted pair group method with arithmetic mean}
\acrodef{WER}   {word error rate}
\acrodef{WPE}   {weighted prediction error}